\documentstyle[psfig,preprint,aps]{revtex}

\tightenlines

\def\beq{\begin{equation}}
\def\eeq{\end{equation}}
\def\bea{\begin{eqnarray}}
\def\eea{\end{eqnarray}}

\begin{document}
\hoffset-1cm
\draft

\title{Dilepton Production from $\rho \>$-Mesons in a Quark-Gluon Plasma
\footnote{Supported by BMBF, GSI Darmstadt, and DFG}}

\author{Markus H. Thoma$^{1,2}$ \footnote{Heisenberg fellow},
Stefan Leupold$^1$, and Ulrich Mosel$^1$}
\address{$^1$Institut f\"ur Theoretische Physik, Universit\"at Giessen, 35392 
Giessen, Germany}
\address{$^2$ECT*, Villa Tambosi, Strada delle Tabarelle 286, 38050 Villazzano
(Trento), Italy}

\date{\today}

\maketitle

\begin{abstract}

Assuming that $\rho\>$-mesons exist in a quark-gluon plasma at temperatures
close to the QCD phase transition, we calculate the dilepton production rate
from $q$-$\bar q$ annihilation via a $\rho\>$-meson state using Vector Meson 
Dominance. The result is compared to the rates from direct $q$-$\bar q$ 
annihilation and from $\pi^+$-$\pi^-$ annihilation. Furthermore we discuss the
suppression of low mass dileptons if the quarks assume an effective mass in 
the quark-gluon plasma.
\end{abstract} 

\bigskip

{\hspace*{1.1cm}Keywords: Relativistic heavy ion collisions, quark-gluon 
plasma,}\\
{\hspace*{1.8cm}dileptons, $\rho$-mesons, Vector Meson Dominance}

\medskip


\narrowtext
\newpage

Dileptons are one of the most promising signatures for medium effects in the 
fireball in relativistic heavy ion collisions.
At SPS there is an 
indication for an enhancement of the dilepton production at invariant masses
between 200 and 800 MeV compared to all known sources neglecting medium
effects \cite{ref2}. Possible explanations of this enhancement are the 
broadening of the width 
of the $\rho\>$-meson by scattering in the medium and the 
reduction of the 
$\rho\>$-mass by the on-set of the restoration of chiral symmetry \cite{ref3}.

Furthermore dileptons may indicate the formation of a quark-gluon plasma (QGP)
phase, in particular at RHIC and LHC \cite{ref4}. 
In a perturbative calculation the lowest order contribution to the 
dilepton production from the QGP comes from the direct $q$-$\bar q$ 
annihilation into a virtual photon (Born term) \cite{ref5}. 
At high invariant masses
$M$ this contribution is believed to dominate. For low invariant masses
of the order of $gT$, on the other hand, $\alpha_s$-corrections become 
important \cite{ref6,ref7,ref8}. 
Close to the critical temperature $g$ might be as 
large as 6 \cite{ref9}. 
Hence $\alpha_s$-corrections could be important even at  
invariant masses up to $M=1$ GeV or larger.

In order to avoid infrared singularities and gauge 
dependent results, HTL resummed propagators and vertices \cite{ref10} have to 
be used in perturbative calculations of the dilepton production from the QGP
\cite{ref7,ref8}.
The dilepton rate follows from the imaginary part of the finite temperature
photon self energy. Unfortunately, it turned out that 2- and probably 
even 3-loop
contributions \cite{ref11} to the photon self energy are of the same order in 
$\alpha_s$ as the 1-loop contribution and exceed the latter one even by 
factors of 3 and more. Thus the application of finite temperature perturbation
theory suffers from the facts that higher order loop corrections are important
even in the weak coupling limit and moreover that realistic values of the 
strong coupling constant are not small. 

The importance of non-perturbative effects in a QGP at temperatures within
reach of heavy ion collisions has been realized in QCD lattice calculations, 
where for
example the equation of state, condensates, and hadronic correlators below and
above the phase transition have been studied \cite{ref12}. 

It has been noticed that the 
equation of state found on the lattice can be described perfectly by an ideal 
gas of massive quarks and gluons (quasiparticles), where an effective,
temperature dependent mass of the order $gT$ for the partons has been 
introduced \cite{ref9}. 
The effective quark and gluon mass at temperatures around the 
critical one $T_c$ is of the order 0.5 GeV. This will lead to a complete
suppression of dileptons from the QGP below about $M=1$ GeV as we will discuss
at the end of the present paper. 

The existence of a gluon condensate above $T_c$ can be used to construct an 
effective quark propagator in the QGP \cite{ref13}. 
The quark dispersion relation following from this
propagator has two branches, of which one shows a minimum at finite momentum.
This leads to an interesting structure of the dilepton spectrum, exhibiting
peaks (Van Hove singularities) and gaps \cite{ref7,ref14}.  

In the present paper we want to study the consequences of the existence of 
hadronic states in the QGP at temperatures around $T_c$ as indicated by
lattice calculations \cite{ref15} 
and the gauged linear sigma model \cite{ref16} on the dilepton production 
from the QGP. 
This temperature regime plays an important role in relativistic heavy ion 
collisons as the mixed phase with $T=T_c$ may exist for a long period
in the fireball and might contribute significantly to the photon and 
dilepton production (see e.g. \cite{ref16a}). In particular, we consider
a possible co-existence of quarks and $\rho\>$-mesonic states in the QGP,
because $\rho\>$-mesons are an important source for dileptons. Although
one expects that $\rho\>$-mesons vanish above $T_c$ quickly with increasing 
temperature \cite{ref15}, they might be present at temperatures around $T_c$
\cite{ref16}.

We assume a simple 
coupling of the quarks to $\rho\>$-mesonic states in the QGP. 
Besides by direct $q$-$\bar q$ annihilation electron-positron pairs can
then be produced by the process $q\bar q\rightarrow \rho
\rightarrow e^+e^-$, where the decay of the $\rho$ will be described using
Vector Meson Dominance (VMD). The result will be compared to the rates 
following from direct annihilation of quarks (Born term) and from 
$\pi^+\pi^-$ annihilation using VMD. Furthermore we assume a vanishing baryon 
density (zero quark chemical potential), equilibrium distributions for the 
quarks and $\rho\>$-mesons, and the bare mass $m_\rho =770$ MeV for the 
$\rho \>$-meson.

For the interaction of quarks with $\rho\>$-mesons we adopt the following
Lagrangian:
\beq 
{\cal L}=-\frac{1}{4}\> \rho_{\mu \nu}^a\> \rho^{\mu \nu}_a+\frac{1}{2}\> 
m_\rho^2\> \rho_\mu ^a \> \rho^\mu _a +\bar q\> \left (i\> \gamma _\mu \> 
\partial ^\mu-m_q+g\> \gamma ^\mu \> \frac{\tau _a}{2}\> \rho_\mu^a\right ) 
q \, ,
\label{lagrangian}
\eeq
where $\rho_{\mu \nu}^a=\partial _\mu \rho _\nu ^a -\partial _\nu \rho _\mu ^a$
with the $\rho\>$-meson field $\rho _\mu^a$ and the quark field $q$. Here 
$a$ denotes the isospin or flavor index and $\tau _a$ the corresponding
isospin matrix. For the quark mass $m_q$ we will investigate two cases,
namely a vanishing bare mass and an effective quark mass $m_q=0.5$ GeV.

To get an idea about the size of the 
coupling constant $g$ describing the strength of the quark-$\rho $ 
coupling we proceed in the following way:
By integrating out the $\rho\>$-mesons from Lagrangian (\ref{lagrangian})
one obtains in lowest order of
the derivative expansion the four-quark interaction term
\begin{equation}
\label{fourquark}
{\cal L}_{\rm 4-quark} = -{1 \over 2} {g^2 \over m_\rho^2} 
\left( \bar q \gamma_\mu {\tau_a \over 2} q \right)^2,
\end{equation}
which might be compared with the corresponding interaction term from
the extended NJL model  $-G_2 \left( \bar q \gamma_\mu \tau_a  
q \right)^2$ \cite{ref17,ref18}.
This suggests the identification $g = \sqrt{8 m_\rho^2 G_2}$
which serves to determine $g$ by taking $G_2$ from the literature 
\cite{ref17,ref18}: $g \approx 5 - 6$. In the following we choose $g=6$.
Of course, it is not clear whether our choice for mass and coupling of a vector 
meson excitation in a hot QGP is justified. However, in lack of any first principle 
calculation we decided to choose values as suggested by measurements or 
calculations for the vacuum case. 

The dilepton production rate (here for massless $e^+$-$e^-$ pairs) can be 
calculated from the imaginary part of the photon self energy according to
\cite{ref19}
\beq
\frac{dN}{d^4xd^4p}=-\frac{\alpha}{12\pi^4}\>\frac{1}{e^{E/T}-1}\>\frac
{Im \Pi_\mu ^\mu (P)}{M^2},
\label{rate}
\eeq
where $\alpha =e^2/4\pi $. Here we use the notation $P=(E,{\vec p}\, )$ and
$p=|{\vec p}\, |$.

Using VMD the photon self energy is related to the $\rho^0\>$-meson 
propagator by
\beq
Im \Pi _\mu ^\mu(P)=\frac{e^2}{g_\rho ^2}\> m_\rho ^4\> Im D_\mu ^\mu (P),
\label{VMD}
\eeq
where $g_\rho=6.07$ \cite{ref19}.
The most general ansatz for the trace of the imaginary part of the full
$\rho\>$-meson propagator reads
\beq
Im D_\mu ^\mu (P)=-\frac{Im F}{(M^2-m_\rho ^2- Re F)^2+(Im F)^2}
-\frac{2\> Im G}{(M^2-m_\rho ^2- Re G)^2+(Im G)^2},
\label{rhoprop}
\eeq 
where $F(p_0,p)$ and $G(p_0,p)$ are the longitudinal and transverse
parts of the $\rho^0$-self energy. To lowest order the $\rho^0$-self energy
is calculated from the one-loop diagram containing a quark loop. It is
given by the one-loop photon self energy containing an electron loop
multiplied by an factor 3/2, where the factor 3 comes from the number of 
quark colors in the loop and the factor 1/2 from the flavor coefficient
${\rm tr}(\tau_0 ^2/4)$, which counts the number of quark flavors 
($u$ and $d$). Furthermore the electron charge $e$ has to be replaced by 
$g$.

The one-loop photon self energy at finite temperature
can be calculated analytically in the high
temperature or equivalently hard thermal loop limit \cite{ref10,ref20}. 
However, in this 
approximation there is no imaginary part for timelike photons, 
$M^2=p_0^2-p^2>0$, resulting in a vanishing dilepton production.
Going beyond the hard thermal loop approximation, 
integral expressions for the 
matter part of the one-loop photon self energy can be derived 
\cite{ref21}: 
\bea
Re F &=& \frac{3}{2\pi ^2}\> g^2\>\frac{M^2}{p^2}\;
\int_0^\infty dk\> k\> n_F(\omega_k) \; \Biggl [
-2\frac{k}{\omega_k} +\frac{4\omega_k^2+M^2}{4p\omega_k}\nonumber \\ 
&&\ln\left |\frac{(2pk+M^2)^2-4p_0^2\omega_k^2}
{(2pk-M^2)^2-4p_0^2\omega_k^2}\right |
+\frac{p_0}{p} \> \ln\left |\frac{M^4-4(p_0\omega_k^2+pk)^2}
{(M^4-4(p_0\omega_k^2-pk)^2}\right |\Biggr ],\nonumber \\
Im F &=& \frac{3}{2\pi}\> g^2\>\frac{M^2}{p^3}\;
\Biggl [\int_{k_-}^{k_+} dk\> k\> n_F(\omega_k)
\left (p_0-\omega_k-\frac{M^2}{4\omega_k}\right )\nonumber \\
&& -2\> p_0\> \theta(-M^2)\; 
\int_{k_-}^{\infty} dk\> k\> n_F(\omega_k) \Biggr ],\nonumber \\
Re G &=& \frac{3}{2\pi ^2}\> g^2\; \int_0^\infty dk\> \frac{k^2}{\omega_k}\> 
n_F(\omega_k) \; \Biggl [
2+\frac{M^2}{p^2}-\left(\frac{\omega_k^2M^2}{2p^3k}+\frac{M^2}{4pk}+
\frac{M^4}{8p^3k}+\frac{m_q^2}{2pk} \right )\> \nonumber \\ 
&&\ln\left |\frac{(2pk+M^2)^2-4p_0^2\omega_k^2}
{(2pk-M^2)^2-4p_0^2\omega_k^2}\right |
-\frac{p_0M^2\omega_k}{2p^3k}\>
\ln\left |\frac{M^4-4(p_0\omega_k^2+pk)^2}
{(M^4-4(p_0\omega_k^2-pk)^2}\right |\Biggr ],\nonumber \\
Im G &=& \frac{3}{4\pi}\> g^2\> \frac{1}{p}\;
\Biggl [\int_{k_-}^{k_+} dk\> k\> n_F(k)
\left (-k+\frac{p_0}{p}\omega_k+\frac{M^2}{2\omega_k}+\frac{M^4}{4p^2\omega_k}
-\frac{m_q^2}{\omega_k}-\frac{p_0M^2}{p^2}
\right )\nonumber \\
&& +\frac{p_0M^2}{p^2}\> \theta(-M^2)\; 
\int_{k_-}^{\infty} dk\> k\> n_F(\omega_k) \Biggr ],
\label{rhoself}
\eea
where $\omega_k^2=k^2+m_q^2$, $n_F(\omega_k)=1/[\exp(\omega_k/T)+1]$, and
\bea
&&k_-=\left |\frac{1}{2}\> \left (p_0\> \sqrt{1-\frac{4m_q^2}{M^2}}-p\right )
\right |,\nonumber \\
&&k_+=\frac{1}{2}\> \left (p_0\> \sqrt{1-\frac{4m_q^2}{M^2}}+p\right ).
\label{kminmax}
\eea

The second term in the imaginary part, being proportional to
$\theta (-M^2)$, describes Landau damping which takes
place only for spacelike momenta, i.e. in scattering processes. Therefore it 
does not contribute to the dilepton production. In the hard thermal loop 
limit, $p_0,p\ll k$, 
the first term of $Im F$ and $Im G$ vanishes, whereas the second 
one reduces to the well known Landau damping contribution in this limit
\cite{ref10,ref20}.

We do not take into account the vacuum part of the $\rho$ self energy because
the decay of a $\rho\>$-meson into a free quark-antiquark pair does not 
take place in the vacuum. The Lagrangian (\ref{lagrangian}) can be 
considered as an effective Lagrangian valid only at temperatures at or above 
the phase transition.

Combining (\ref{rate}) to (\ref{rhoself}) we obtain the dilepton 
production rate
($\rho$-quark rate) by numerical integration over the loop momentum $k$ 
(magnitude of the three-momentum). 
In Fig.1 this rate is compared to the Born rate 
(direct quark-antiquark annihilation) \cite{ref5} 
and the $\pi^+$-$\pi^-$ annihilation
rate via VMD ($\rho$-$\pi$ rate) calculated by Gale and Kapusta \cite{ref19}.
In contrast
to the Born rate the $\rho$-quark 
rate as well as the $\rho$-$\pi$ rate show 
a clear peak in the vicinity of the $\rho\>$-mass. For $M<0.4$ GeV the
$\rho$-quark 
rate agrees well with the Born rate, which both have no mass cut 
assuming vanishing $u$- and $d$-quark masses. For $M>0.4$ GeV the $\rho$-quark 
rate agrees better with the $\rho$-$\pi$ rate, which is zero for
$M<0.28$ GeV due to the finite $\pi$-mass. The agreement of the absolute
values of these rates might be accidental
as, for example, the $\rho$-quark rate 
is enhanced compared to the $\rho$-$\pi$
rate by the larger number of quark  
degrees of freedom than $\pi$ degrees of freedom, but reduced
by the neglect of vacuum contributions in the first one.

We have chosen a temperature
of $T=0.15$ GeV in accordance with the critical temperature 
for the QCD phase transition as predicted by QCD lattice theory, since we 
expect $\rho\>$-mesons to co-exist with quarks only close to $T_c$. For the
momentum of the virtual photon we have chosen $p=0.2$ GeV, because CERES-SPS
data show that the dilepton enhancement occurs at small transverse
momenta $0.2$ GeV$<p_T<0.5$ GeV \cite{ref2}.

In Fig.2 the $\rho $-quark rate is shown for different temperatures ($T=0.15$
GeV and $T=0.2$ GeV). In Fig.3 the $p$-dependence of this rate is presented 
by comparing the rate at $p=0.2$ GeV and $p=1$ GeV.

Finally, we discuss the possibility of an effective, temperature dependent 
quark mass as indicated by comparing the equation of state of an ideal gas 
of massive quarks and gluons with lattice calculations. In the temperature
regime between $T_c$ and about $2T_c$, quark masses of the order of $m=500$
GeV are required to match the quasiparticle equation of state to lattice
results \cite{ref9}. 
In Fig.4 the influence of such an effective quark mass on the Born 
rate is shown. The rate vanishes for $M<2m_q$
but approaches quickly the bare mass rate above $M=1$ GeV. 
The same behavior can be observed in the $\rho$-quark rate (Fig.5),
where the $\rho\>$-peak now is absent in the case of finite
quark masses.

If this simple picture for the QGP were true, 
there would be no dilepton 
production from the QGP below the $\rho\>$-peak, where the SPS dilepton 
enhancement has been observed. On the other hand, although the simple 
assumption of an effective quark mass might be sufficient to explain the
equation of state, the correct quark dispersion relation in a QGP may look 
completely different. Perturbative as well as non-perturbative approaches
indicate two quark branches starting from the same effective mass at zero 
momentum. The lower branch, 
corresponding to a collective quark mode (plasmino)
that is absent in vacuum, 
shows a minimum at finite momentum. The splitting of 
the two collective quark modes and the minimum of the plasmino branch give 
rise to interesting structures, namely peaks (van Hove singularities)
and gaps, in the dilepton rate. In particular, low mass dileptons are possible
coming from electromagnetic transitions from the upper to the lower
branch \cite{ref7,ref14}. Furthermore, a finite width of the quarks from 
scattering in the QGP will change the dilepton production rate
in addition.

Summarizing, we have calculated the dilepton production rate from 
quark-antiquark annihilation assuming the co-existence of quarks and 
$\rho\>$-mesons around the QCD phase transition as indicated by lattice QCD 
and gauged linear sigma model calculations. Using VMD the dilepton
production rate has been calculated in a similar way as the
rate from $\pi^+$-$\pi^-$ rate, replacing the $\pi$-loop by a quark loop in 
the latter. For low invariant masses the rate turns out to agree well with
the Born rate, whereas for higher invariant masses the agreement with the
$\rho$-$\pi$ rate is better. Although the coincidence in the absolute 
value of these rates may be accidental, the agreement of the shape is easy 
to understand (no mass cut for massless quarks, $\rho\>$-peak in VMD). 

Concerning SPS data the contribution from the $\rho$-quark rate to the 
dilepton spectrum is by far too small to explain the observed enhancement
\cite{ref2}.
For the $\rho$-quark rate is of the same order as the Born rate, which
leads to a dilepton yield that is about two orders of magnitude smaller than 
the observed one according to hydrodynamical calculations \cite{ref16a}. 
However, this situation will be different at RHIC and LHC where a higher
initial temperature and a longer lifetime of the QGP are expected.

Finally,
we have shown that an effective quark mass as indicated by lattice 
calculations of the equation of state of the QGP leads to a suppression
of dileptons with invariant masses below about 1 GeV. However, the simple 
picture of non-interacting massive quarks is probably too oversimplified
for predicting the dilepton production from the QGP.



\begin{figure}

\centerline{\psfig{figure=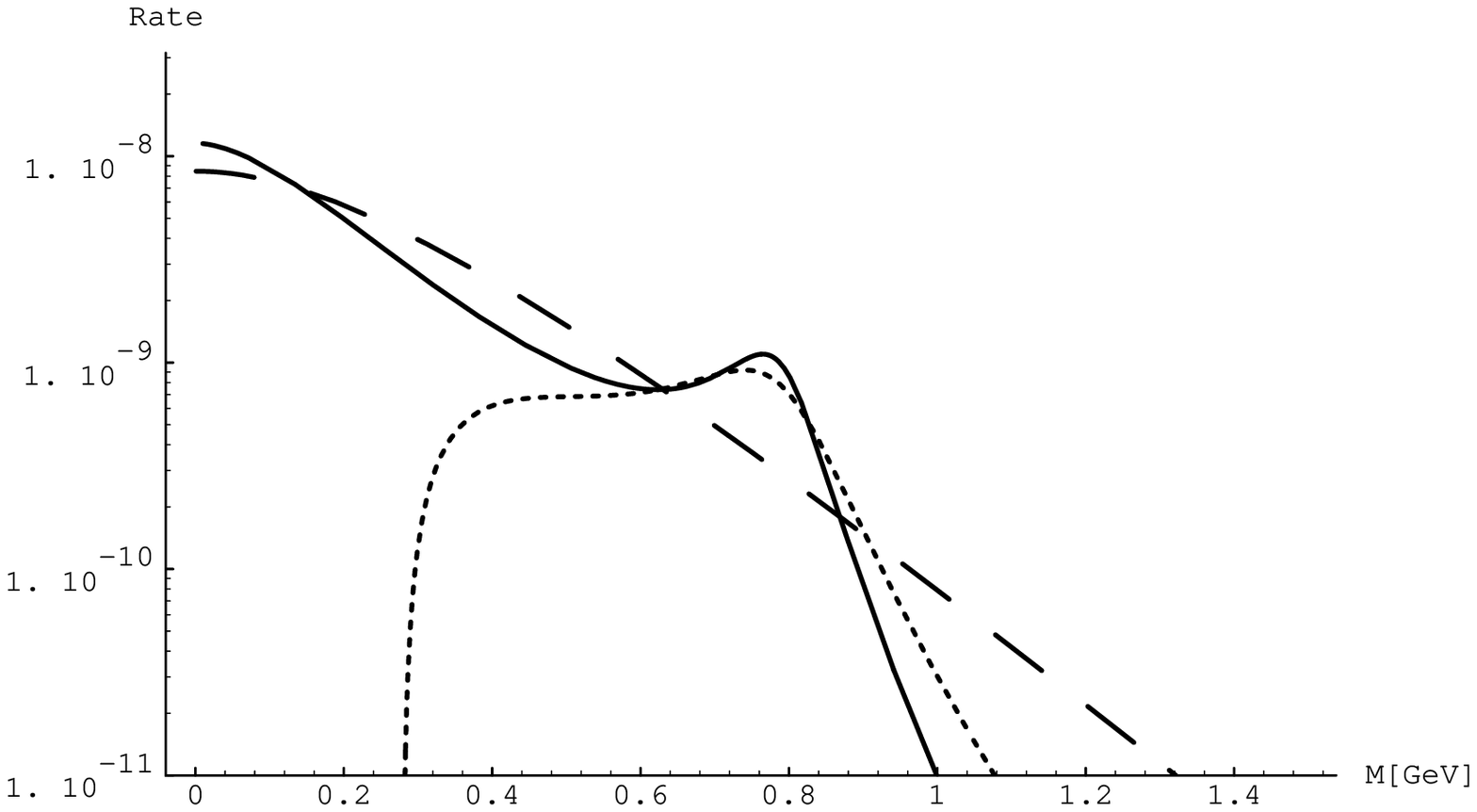,width=7.5cm}}
\caption{$\rho$-quark rate (solid line), Born rate (dashed line), and 
$\rho$-$\pi$ rate (dotted line) at $T=0.15$ GeV, $p=0.2$ GeV and $m_q=0$.}

\end{figure}

\vspace*{-1cm}

\begin{figure}

\centerline{\psfig{figure=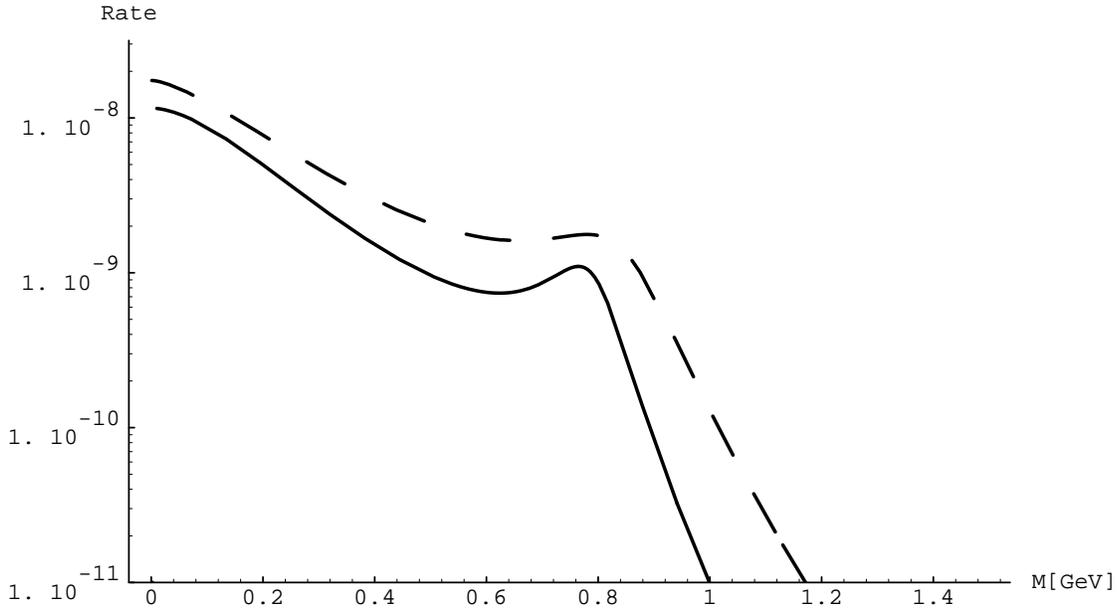,width=7.5cm}}
\caption{$\rho$-quark rate at $T=0.15$ GeV (solid line), $T=0.2$ GeV (dashed 
line), $p=0.2$ GeV, and $m_q=0$.}

\end{figure}

\begin{figure}

\centerline{\psfig{figure=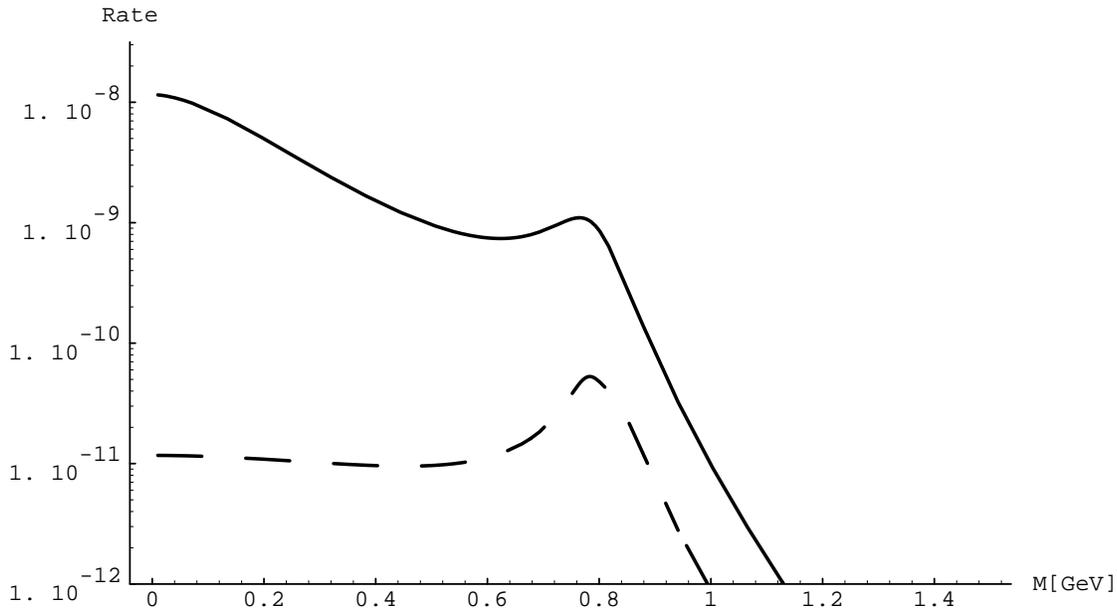,width=7.5cm}}
\caption{$\rho$-quark rate at $T=0.15$ GeV, $p=0.2$ GeV (solid line),
$p=1$ GeV (dashed line), and $m_q=0$.}

\end{figure}

\vspace*{-1cm}

\begin{figure}

\centerline{\psfig{figure=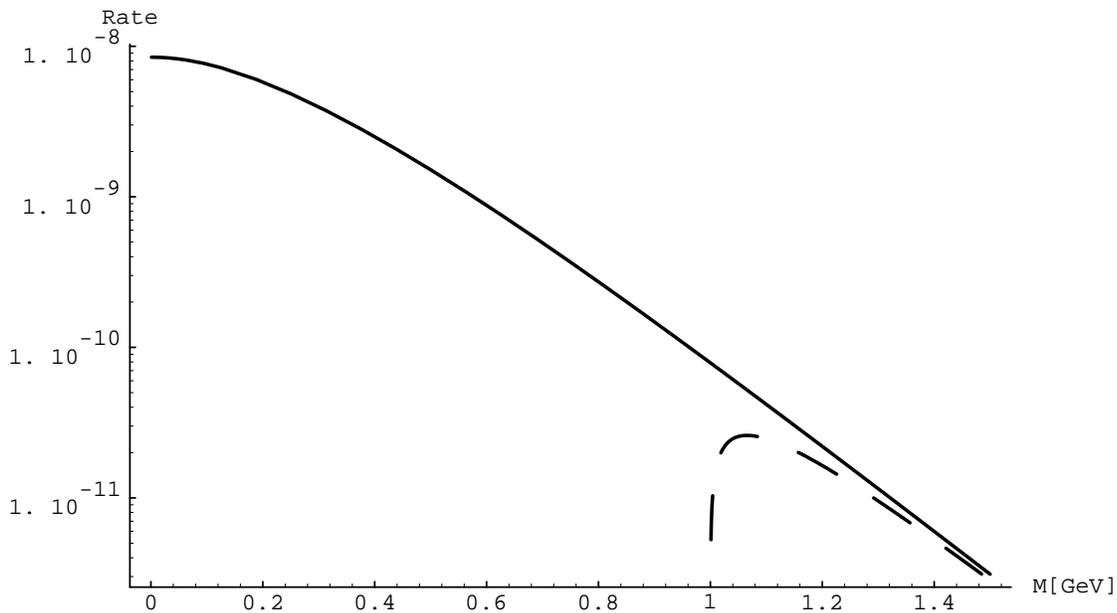,width=7.5cm}}
\caption{Born rate at $T=0.15$ GeV, $p=0.2$ GeV, $m_q=0$ (solid line),
and $m_q=0.5$ GeV (dashed line).}

\end{figure}

\begin{figure}

\centerline{\psfig{figure=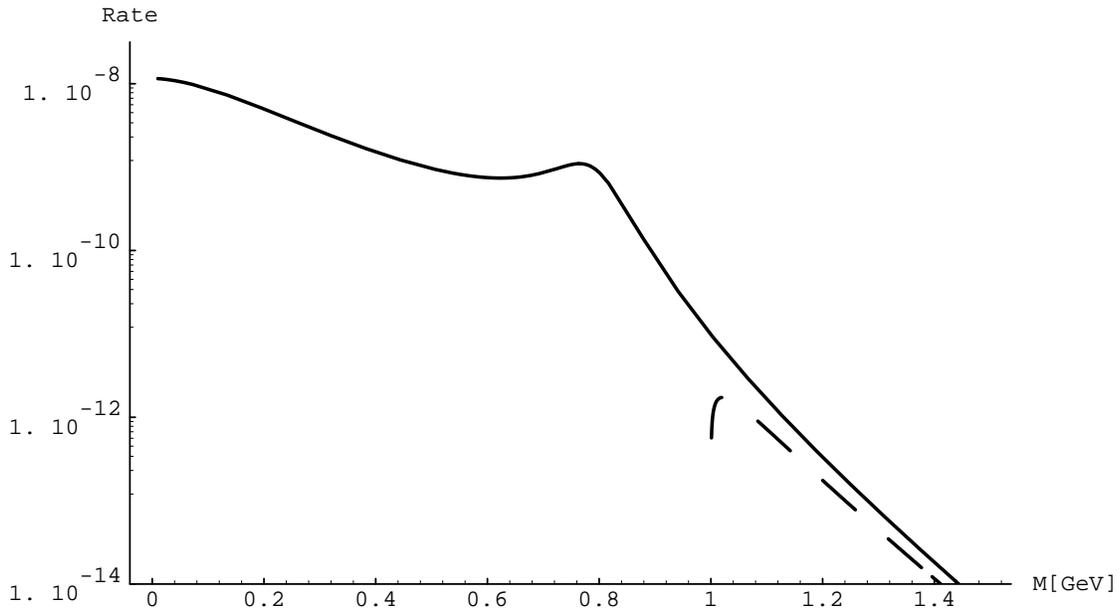,width=7.5cm}}
\caption{$\rho$-quark rate at $T=0.15$ GeV, $p=0.2$ GeV, $m_q=0$ (solid line),
and $m_q=0.5$ GeV (dashed line).}

\end{figure}

\end{document}